\pdfoutput=1

\documentclass[pra,aps,superscriptaddress,amsmath,twocolumn,amssymb,nofootinbib]{revtex4-1}

\usepackage{braket,verbatim,bm,bbold,amsmath,amssymb,epsfig,float,color}
\usepackage[colorlinks,bookmarks=false,citecolor=red,linkcolor=blue,hyperfootnotes=true,urlcolor=magenta]{hyperref}
\usepackage{color}
\usepackage[dvipsnames]{xcolor}
\usepackage{enumitem}   
\usepackage{amsmath,amsfonts,amssymb,stackengine,graphicx}
\setstackgap{S}{-1.5pt}

\newcommand{\nc}{\newcommand}
\nc{\ir}{\mathrm{i}}
\nc{\dd}{\mathrm{d}} 
\nc{\eE}{\mathsf{e}}
\nc{\Tr}{\text{Tr}}
\nc{\id}{\mathbb{I}}
\nc{\I}{\mathcal{I}}
\nc{\nnF}{\mathcal{F}}

\begin{document}

\title{Symmetry-resolved R\'enyi fidelities and quantum phase transitions }  

\author{Gilles Parez}
\email{gilles.parez@umontreal.ca}
\affiliation{\it Centre de Recherches Math\'ematiques (CRM), Universit\'e de Montr\'eal, P.O. Box 6128, Centre-ville
Station, Montr\'eal (Qu\'ebec), H3C 3J7, Canada}
\date{\today}

\begin{abstract}
We introduce a family of quantum R\'enyi fidelities and discuss their symmetry resolution. We express the symmetry-resolved fidelities as Fourier transforms of charged fidelities, for which we derive exact formulas for fermionic Gaussian states. These results also yield a formula for the total fidelities of fermionic Gaussian states, which we expect to have applications beyond the scope of this paper. We investigate the total and symmetry-resolved fidelities in the XX spin chain, and focus on (i) fidelities between thermal states, and (ii) fidelities between reduced density matrices at zero temperature. Both thermal and reduced fidelities can detect the quantum phase transition of the XX spin chain. Moreover, we argue that symmetry-resolved fidelities are sensitive to the inner structure of the states. In particular, they can detect the phase transition through the reorganization of the charge sectors at the critical point. This a main feature of symmetry-resolved fidelities which we expect to be general. We also highlight that reduced fidelities can detect quantum phase transitions in the thermodynamic limit. 
\end{abstract}

\maketitle
\section{Introduction}\label{sec:intro}

Quantum phase transitions (QPTs) are ubiquitous in statistical and condensed matter physics \cite{S07}. A QPT is characterized by a strong reorganization of the groundstate $|\psi(\lambda)\rangle$ of a quantum many-body Hamiltonian $H(\lambda)$ as an external parameter crosses a critical value $\lambda=\lambda_c$. In contrast with thermal phase transitions, QPTs are not driven by temperature. Instead, the parameter $\lambda$ is typically an interaction strength, a magnetic field or an anisotropy parameter. 

Over the last two decades, there has been an important cross fertilization between quantum information and condensed matter physics. In particular, it has proven very efficient to use quantities originally defined in the context of quantum information, such as entanglement measures~\cite{VPRK97} and quantum fidelities \cite{U76,J94}, to detect and characterize QPTs in many-body systems \cite{OAFF02, ON02, VLRK03, CC04,QSLZS06,zanardi2006ground,cozzini2007quantum,ZB08}. Since these breakthroughs, the investigation of entanglement measures has become a prominent research area in condensed matter and statistical physics \cite{amico2008entanglement,calabrese2009entanglement,laflorencie2016quantum,calabrese2016introduction}. Quantum fidelities have generated interest \cite{G10}, but they did not receive a treatment as thorough and systematic as entanglement entropies and related measures did in the context of quantum many-body systems. 

Quantum fidelities and entanglement measures share important properties. For instance, in one-dimensional quantum critical systems, the so-called (logarithmic) bipartite fidelity~\cite{DS11,SD13} exhibits a logarithmic violation of the area law and encodes conformal data of the underlying conformal field theory (CFT) \cite{DS11,SD13,hagendorf2017open,PMDR19,MDPL21,HP21}, similarly to the entanglement entropy~\cite{CC04}. Moreover, fidelities are able to detect topological phase transitions \cite{hamma2008entanglement,abasto2008fidelity,eriksson2009reduced}, and the bipartite fidelity possess a topological term \cite{stephan2012renyi,stephan2013entanglement} that is related to the topological entanglement entropy~\cite{kitaev2006topological,levin2006detecting}.

A quantum fidelity $F(\rho,\sigma)$ measures the similarity between two density matrices $\rho$ and $\sigma$. If $\rho$ and $\sigma$ are projectors on the normalized pure states $|\psi_\rho\rangle$ and $| \psi_\sigma\rangle$, respectively, a natural measure of similarity between these states is the overlap $|\langle \psi_\rho | \psi_\sigma \rangle|$. We interpret this overlap as the so-called \textit{pure-state fidelity} between the operators $\rho = |\psi_\rho \rangle \langle \psi_\rho|$ and  $\sigma = |\psi_\sigma \rangle \langle \psi_\sigma|$. 

Let us review how pure-state fidelities are suited to detect QPTs \cite{zanardi2006ground}. We consider the fidelity 
\begin{equation}
\label{eq:Fpure}
F_{\textrm{pure}}(\lambda,\dd \lambda)=|\langle \psi(\lambda-\dd \lambda)|\psi(\lambda+\dd \lambda)\rangle|
\end{equation}
between two groundstates of a many-body Hamiltonian $H(\lambda)$ that possesses a QPT at $\lambda=\lambda_c$. For finite sizes and away from the critical point, we expect $F(\lambda,\dd \lambda) \sim 1$ for small $\dd \lambda$, because the two states are nearly identical. In stark contrast, at the critical point, because of the strong reorganization of the groundstate, the two states become orthogonal and $F(\lambda_c,\dd \lambda) \sim 0$. In the thermodynamic limit, the groundstates at different values of $\lambda$ become mutually orthogonal, so that $F(\lambda,\dd \lambda)$ vanishes for all values of $\dd \lambda \neq 0$ and $\lambda$. However, in the vicinity of the critical point, this suppression of the fidelity is detectable for finite sizes. 

In the more general case where $\rho$ and/or $\sigma$ are mixed density matrices, we say that the resulting fidelity is a \textit{mixed-state fidelity}. The most common definition of mixed-state fidelity is the so-called Uhlmann-Jozsa fidelity \cite{U76,J94}:
\begin{equation}
\label{eq:UJF}
F(\rho,\sigma) = \Tr \{(\sqrt{\rho}\sigma \sqrt{\rho})^{1/2}\}.
\end{equation}
This quantity satisfies a set of properties \cite{J94,G10} that we discuss below. In particular, $F(\rho,\sigma)$ reduces to the pure-state fidelity $|\langle \psi_\rho | \psi_\sigma \rangle|$ when both operators project on a pure state, and $0\leqslant F(\rho,\sigma) \leqslant 1$ with $F(\rho,\sigma)=1 $ if and only if $\rho=\sigma$. 

We mention two physical contexts in which the use of mixed-state fidelity is relevant. First, we consider a system at finite temperature. Because of thermal fluctuations, the system is described by the thermal density matrix $\rho(\beta,\lambda) = \eE^{-\beta H(\lambda)}/\Tr(\eE^{-\beta H(\lambda)})$, where $\beta$ is the inverse temperature. A fidelity between two thermal density matrices is called a \textit{thermal fidelity}. A striking feature of thermal fidelities is that they allow us to detect the zero-temperature QPT of the underlying Hamiltonian $H(\lambda)$, even at finite temperature $\beta < \infty$ \cite{zanardi2007mixed}. Second, we consider a system at zero temperature in the groundstate $|\psi(\lambda)\rangle$ of $H(\lambda)$, and assume that we have access only to a subsystem $A$ of the whole bipartite system $A\cup B$. In that case, the system $A$ is described by the reduced density matrix $\rho_A = \Tr_B(|\psi(\lambda)\rangle \langle \psi(\lambda)|)$, and the fidelity between two reduced density matrices is called a \textit{reduced fidelity}~\cite{paunkovic2008fidelity}. Reduced fidelities are also known to detect QPTs \cite{paunkovic2008fidelity,ma2008reduced,kwok2008partial,eriksson2009reduced,xiong2009reduced,ma2009many,wang2010reduced}. However, in most examples discussed in the literature, the subsystem $A$ consists of a small number of degrees of freedom (typically one or two sites for lattice models). In Refs \cite{ma2008reduced,ma2009many}, the authors investigate reduced fidelities for larger systems in the special case where the reduced density matrix has a $2 \times 2$ block-diagonal form. Hence, to the best of our knowledge, reduced fidelities have not been investigated in generic cases where $A$ is a genuine macroscopic subregion with a large number of degrees of freedom. 

In the context of a quantum many-body system with a global conserved charge, an important timely challenge is to understand how the various symmetry sectors contribute to the total entanglement. This so-called \textit{symmetry resolution of entanglement} \cite{lr-14,GS18,equi-sierra} has been investigated in the contexts of critical systems \cite{barghathi2018renyi,barghathi2019operationally,bons,bons-20,trac-20,mrc-20,eimd-21,jones2022symmetry,ares2022symmetry,ares2022multi}, integrable systems and field theories \cite{mdc-20,mdc-20b,hc-20,hcc-21,capizzi2022symmetry,capizzi2022symmetrybis}, topological phases \cite{as-20,Topology,oblak2022equipartition} and non-equilibrium quantum many-body systems~\cite{fg-19,fg-21,PBC21,PBC21bis,PBC22,scopa2022exact,piroli2022thermodynamic,chen2022dynamics}. Many entanglement-related quantities have since acquired their symmetry-resolved counterpart, such as the mutual information \cite{PBC21bis,ares2022multi}, entanglement negativity \cite{cgs-18,mbc-21,PBC22}, relative entropies and trace distance \cite{cc-21,c-21}, the Page curve~\cite{murciano2022symmetry} and the entanglement asymmetry \cite{ares2022entanglement}. However, a symmetry-resolved version of quantum fidelities is still lacking. An intriguing question is to understand how each symmetry sector contributes to the total fidelities close to a QPT, as well as in the critical regime. In this paper, we introduce a family of quantum fidelities, that we dub R\'enyi fidelities, and investigate their symmetry resolution in the vicinity of a QPT. Since the symmetry resolution of pure-state fidelity is trivial, we focus on thermal and reduced ones. We choose to study these quantities in the XX spin chain \cite{lieb1961two}. This model is the perfect test bed for these investigations, because it has a global $U(1)$ symmetry, exhibits a QPT, and is exactly solvable.

This paper is organized as follows. In Sec. \ref{sec:def} we define the R\'enyi fidelities and show that they obey natural generalizations of the axioms of the Uhlmann-Jozsa fidelity. We also define the symmetry-resolved fidelities and express them as the Fourier transforms of charged fidelities. We conclude the section with formulas for the charged and total fidelities of fermionic Gaussian states. In Sec. \ref{sec:XX}, we introduce the XX spin chain in an external magnetic field $h$ and discuss its QPT at $h_c=1$. As a first application, we investigate the total and symmetry-resolved thermal fidelities in Sec. \ref{sec:thermal}, where the diagonal form of the XX Hamiltonian allows us to provide a large number of exact results and approximations that we verify numerically. We study the reduced fidelities of the XX spin chain in the thermodynamic limit in Sec.~\ref{sec:red} and solely rely on numerical evaluations of the formulas for Gaussian states. We observe similar qualitative behavior as for the thermal fidelities, and in particular the reduced fidelities are able to detect the QPT at $h=1$. This result shows that reduced fidelities are able to detect QPTs in the thermodynamic limit, which is something that pure-state fidelities fail to do \cite{zanardi2006ground}. Finally, we conclude with a discussion of the main results and future potential research directions in Sec. \ref{sec:ccl}.

\section{Quantum fidelities}\label{sec:def}

\subsection{Definition}

In this section, we introduce a family of quantum fidelities that generalize the Uhlmann-Jozsa fidelity, see Eq.~\eqref{eq:UJF}. For two density matrices $\rho, \sigma$, we define the R\'enyi fidelity of index $n$ as
\begin{equation}\label{eq:Ftot}
F_n(\rho, \sigma) = \frac{\Tr\{(\rho \sigma)^n\}}{\sqrt{\Tr(\rho^{2n})\Tr(\sigma^{2n})}}. 
\end{equation}
In the following, we restrict our attention to density matrices that are positive semi-definite and have unit trace. We note that the R\'enyi fidelity reduces to the Uhlmann-Jozsa fidelity for $n = 1/2$ if the density matrices commute.\footnote{One could instead define a double-index R\'enyi fidelity as $$F_{n,m}(\rho,\sigma) = \frac{\Tr\{(\rho^m \sigma^{2m}\rho^m)^n\}}{\sqrt{\Tr(\rho^{4nm})\Tr(\sigma^{4nm})}}.$$ This quantity coincides with the Uhlmann-Jozsa fidelity in the limit $n,m \to 1/2$, irrespective of the commutation relation between $\rho$ and $\sigma$. However, the forthcoming calculations for $F_n(\rho)$ generalize to these double-index R\'enyi fidelities, and therefore we focus on $F_{n}(\rho,\sigma)$. Understanding the qualitative differences between $F_{n}(\rho,\sigma)$ and $F_{n,m}(\rho,\sigma)$ in the case of non-commuting density matrices constitutes an interesting future direction that will be addressed in subsequent work.} Moreover, for $n=1$ we recover the fidelity introduced in Ref. \cite{WYY08}. 

The R\'enyi fidelity $F_n(\rho, \sigma)$ satisfies the following properties for $n>0$:
\begin{subequations}
\begin{enumerate}[label=(\roman*)]
\item Normalization: \begin{equation}
0 \leqslant F_n(\rho, \sigma)\leqslant 1, \qquad F_n(\rho, \sigma)= 1 \Leftrightarrow \rho=\sigma,
\end{equation}
\item Symmetry:
\begin{equation}
F_n(\rho,\sigma)=F_n(\sigma,\rho),
\end{equation}
\item Invariance under unitary transformation $U$: 
\begin{equation}
F_n (U\rho U^{-1},U\sigma U^{-1})=F_n(\rho, \sigma),
\end{equation}
\item Multiplicativity: 
\begin{equation}
F_n(\rho_1\otimes \rho_2,\sigma_1\otimes \sigma_2)=F_n(\rho_1,\sigma_1)F_n(\rho_2,\sigma_2),
\end{equation}
\item Simplification for pure states: 
\begin{equation}
\begin{split}
F_n(\rho, |\psi_{\sigma}\rangle \langle \psi_{\sigma}| ) &= \frac{\langle \psi_\sigma | \rho|\psi_\sigma \rangle^n}{\sqrt{\Tr(\rho^{2n})}}, \\[.3cm]
F_n(|\psi_{\rho}\rangle \langle \psi_{\rho}|, |\psi_{\sigma}\rangle \langle \psi_{\sigma}|)&=| \langle \psi_{\rho}| \psi_{\sigma} \rangle |^{2n}.
\end{split}
\end{equation}
\end{enumerate}
\end{subequations}
These are natural generalizations of the properties satisfied by the Uhlmann-Jozsa fidelity \cite{J94,G10}. 

\subsection{Symmetry resolution}

Let us consider the case where the density matrices commute with a $U(1)$ charge $Q$, $[\rho,Q]=[\sigma,Q]=0$. Both density matrices thus have a block diagonal form,
\begin{equation}
\label{eq:mat}
\rho = \oplus_q p_\rho(q) \rho(q), \quad \sigma = \oplus_q p_\sigma(q) \sigma(q), 
%
\end{equation}
where $q$ are the eigenvalues of $Q$, and $p_{\rho}(q), p_{\sigma}(q)$ are the probabilities of measuring the charge~$q$ in the states $\rho$ and $\sigma$, respectively. The probabilities satisfy the normalization condition $\sum_q p_\rho(q) = \sum_q p_\sigma(q)=1$. Moreover, the density matrices $\rho(q),\sigma(q)$ have trace one and are interpreted as density matrices in the charge sector $q$. We define the \textit{symmetry-resolved R\'enyi fidelities} $F_n(\rho,\sigma;q)$ as 
\begin{equation}
\label{eq:SRF_def}
\begin{split}
F_n(\rho,\sigma;q)  &\equiv F_n(\rho(q),\sigma(q))\\
&= \frac{\Tr\{(\rho(q) \sigma(q))^n\}}{\sqrt{\Tr(\rho(q)^{2n})\Tr(\sigma(q)^{2n})}}.
\end{split}
\end{equation}

We mention that the symmetry resolution of pure-state fidelities is trivial. Indeed, if $[\rho,Q]=0$ and $\rho = |\psi_\rho\rangle \langle \psi_\rho|$, this implies that $Q |\psi_\rho \rangle = q_\rho |\psi_\rho \rangle$, and similarly for $\sigma$. Hence, both density matrices have only one symmetry sector, and $F_n(\rho,\sigma ;q) = \delta_{q,q_\rho}\delta_{q_\rho,q_\sigma} |\langle \psi_\rho |\psi_\sigma \rangle |^{2n}$. In the following we thus focus on mixed-state symmetry-resolved fidelities. 

To proceed, we introduce the non-normalized symmetry-resolved fidelities $\nnF_n(\rho,\sigma;q)$,
\begin{equation}
\nnF_n(\rho,\sigma;q) \equiv( p_\rho(q) p_\sigma(q)) ^n  \frac{\Tr\{(\rho(q) \sigma(q))^n\}}{\sqrt{\Tr(\rho^{2n})\Tr(\sigma^{2n})}}.
\end{equation}

There is a simple relation between the total fidelities and the non-normalized symmetry-resolved ones:
\begin{equation}
\label{eq:Ftotq}
\begin{split}
F_n(\rho,\sigma)
& =\sum_q\nnF_n(\rho,\sigma;q).
\end{split}
\end{equation}

In the rest of the paper, we study the non-normalized symmetry-resolved fidelities $\nnF_n(\rho,\sigma;q)$ instead of $F_n(\rho,\sigma;q)$, and thus refer to  $\nnF_n(\rho,\sigma;q)$ as the symmetry-resolved fidelities. The reason is the following. If the probabilities are zero or very small, the fidelities $F_n(\rho,\sigma;q)$ may nonetheless be non-vanishing and amount to comparing matrices essentially filled with zeros. In contrast, $\nnF_n(\rho,\sigma;q)$ is zero if at least one probability vanishes, and it reflects the true symmetry decomposition of the total fidelities, as highlighted in Eq.~\eqref{eq:Ftotq}.

\subsection{Charged fidelities and Fourier transform}

From the total density matrices, it is in general a hard problem to extract the contribution of a symmetry sector, because it requires the knowledge of the full symmetry decomposition of the matrices. A way to circumvent this issue, proposed in Ref.~\cite{GS18}, is to express symmetry-resolved quantities as the Fourier transform of certain charged moments. Following this idea, we introduce the charged R\'enyi fidelities $f_n(\rho,\sigma;\alpha)$ as
\begin{equation}
\label{eq:fnDef}
f_n(\rho,\sigma;\alpha) = \frac{\Tr\{(\rho \sigma)^n \eE^{\ir \alpha Q}\}}{\sqrt{\Tr(\rho^{2n})\Tr(\sigma^{2n})}}.
\end{equation}
They satisfy $f_n(\rho,\sigma;\alpha=0)=F_n(\rho,\sigma)$, and their Fourier transform yields the symmetry-resolved fidelities\footnote{To compute $F_n(\rho,\sigma;q)$ defined in Eq. \eqref{eq:SRF_def} with this approach, one needs to introduce the charged moments $Z_n(\chi; \alpha) = \Tr(\chi^n \eE^{\ir \alpha Q})$ with $\chi=\rho, \sigma, \rho\sigma$, and compute the Fourier transform of these three quantities separately. Since we do not investigate $F_n(\rho,\sigma;q)$ in this paper and these techniques are now standard in the literature, we do not discuss them further. We refer the interested reader to Ref. \cite{bons} for a pedagogical introduction to the technicalities of symmetry-resolved entanglement computations.}, 
\begin{equation}
\label{eq:FnqInt}
\nnF_n(\rho,\sigma;q)=\int_{-\pi}^{\pi} \frac{\dd \alpha}{2 \pi}\eE^{-\ir \alpha q}f_n(\rho,\sigma;\alpha).
\end{equation}

\subsection{Formulas for fermionic Gaussian states}  \label{sec:Gaussian}

Let us now assume that $\rho$ and $\sigma$ are fermionic Gaussian operators. In that case, the spectrum of both matrices is obtained from the related two-point correlation matrices $C_{\rho}$ and $C_{\sigma}$ \cite{chung2001density,peschel2003calculation,peschel2009reduced}. Their matrix elements are
\begin{equation}
[C_{\rho}]_{j,k} = \Tr(\rho c_j^\dagger c_k), \quad [C_{\sigma}]_{j,k} = \Tr(\sigma c_j^\dagger c_k),
\end{equation}
where $c_j^\dagger,c_k,$ are fermionic creation and annihilation operators that satisfy the usual anticommutation relations 
\begin{equation}
\label{eq:anticom}
\{c_j^\dagger,c_k\}=\delta_{j,k}, \quad \{c_j,c_k\}=\{c_j^\dagger,c_k^\dagger\}=0.
\end{equation}

We introduce the matrices 
\begin{equation}
J_{\rho} = 2C_{\rho}-\mathbb{I}, \quad J_{\sigma} = 2C_{\sigma}-\mathbb{I},
\end{equation}
and
\begin{equation}
\label{eq:Jbull}
J_{\bullet} = (\mathbb{I}+J_{\rho}J_{\sigma})^{-1} (J_{\rho}+J_{\sigma}),
\end{equation}
and denote their respective eigenvalues by $\nu_j^{\rho}, \nu_j^{\sigma},\nu_j^{\bullet}$. With standard techniques of Gaussian operators and Refs. \cite{fagotti2010entanglement,eisert2018entanglement}, we find
\begin{widetext}
\begin{equation}
\label{eq:fnGauss}
f_n(\rho,\sigma;\alpha) =  \left[\det \left(\frac{\mathbb{I}+J_{\rho}J_{\sigma}}{2}\right)\right]^n \prod_j \frac{ \left[\left(\frac{1+\nu_j^{\bullet}}{2}\right)^n \eE^{\ir \alpha}+ \left(\frac{1-\nu_j^{\bullet}}{2}\right)^n\right]}{  \left[\left(\frac{1+\nu_j^{\rho}}{2}\right)^{2n} + \left(\frac{1-\nu_j^{\rho}}{2}\right)^{2n}\right]^{1/2} \left[\left(\frac{1+\nu_j^{\sigma}}{2}\right)^{2n} + \left(\frac{1-\nu_j^{\sigma}}{2}\right)^{2n}\right]^{1/2}}.
\end{equation}
\end{widetext}

In the case where $[\rho,\sigma]=0$, it is not necessary to introduce the matrix $J_\bullet$, and the numerator of Eq. \eqref{eq:fnGauss} simplifies to
 $$\prod_j\left[\left(\frac{(1+\nu_j^{\rho})(1+\nu_j^{\sigma})}{4}\right)^n \eE^{\ir \alpha}+ \left(\frac{(1-\nu_j^{\rho})(1-\nu_j^{\sigma})}{4}\right)^n\right].$$
 Moreover, in that case, the definition of Eq. \eqref{eq:fnDef} implies $f_n(\rho,\sigma;\alpha) =f_1(\rho^n,\sigma^n;\alpha) $, and hence the $n$ dependence is trivial.

We stress that the limit $\alpha \to 0$ of Eq.~\eqref{eq:fnGauss} gives a formula for the total fidelities of generic and non-commuting Gaussian states. We believe that Eq.~\eqref{eq:fnGauss} will enable the systematic investigation of mixed-state fidelities in the context of quantum many-body systems, both in and out of equilibrium. Moreover, the limit $n \to 1/2$ is direct, and hence Eq.~\eqref{eq:fnGauss} yields an exact formula for the Uhlmann-Jozsa fidelity. This provides an additional motivation for the investigation of the R\'enyi fidelities. As discussed in the footnote below Eq.~\eqref{eq:Ftot}, it is also possible to define a double-index R\'enyi fidelity $F_{n,m}(\rho,\sigma)$, and the generalization of Eq.~\eqref{eq:fnGauss}  to the double-index case is direct. For these reasons, we highlight Eq.~\eqref{eq:fnGauss} as a main result of this paper.

\section{XX spin chain and QPT} \label{sec:XX}

\subsection{Definition}
In this section, we review the spin-$1/2$ XX spin chain with periodic boundary conditions in a magnetic field~\cite{lieb1961two}. The Hamiltonian is 
\begin{equation}
H(h)=-\frac{1}{4}\sum_{j=1}^N (\sigma_{j+1}^x\sigma_{j}^x+\sigma_{j+1}^y\sigma_{j}^y+2h \sigma_j^z),
\end{equation}
where $N$ is the length of the chain, $h$ is the magnetic field, and $\sigma_j^a$, $a=x,y,z$, is the Pauli matrix $\sigma^a$ acting on site~$j$. There is a symmetry between $\pm h$, and we focus on $h\geqslant 0$. For simplicity, we also assume $N$ to be an even number. This Hamiltonian is equivalent to a tight-biding Hamiltonian through the Jordan-Wigner transformation, and its diagonalization is standard. In the large-$N$ limit, the Hamiltonian in diagonal form is 
\begin{equation}
\label{eq:HDiag}
H(h)=\sum_{k=1}^N \epsilon_k d_k^\dagger d_k, \quad \epsilon_k = -\cos\left(\frac{2 \pi k}{N}\right)+h,
\end{equation}
were $d_k^\dagger,d_k$ are Fourier transforms of the fermion operators $c_j^\dagger,c_j$, and they satisfy the fermionic anticommutation relations of Eq.~\eqref{eq:anticom}. 

The XX Hamiltonian chain possesses a $U(1)$ symmetry with global charge 
\begin{equation}
Q = \sum_{k=1}^N c_k^\dagger c_k.
\end{equation}
Physically, $H(h)$ conserves the number of fermions, or the magnetization in the spin-chain picture. 

For $0 \leqslant h\leqslant1$, the groundstate of the Hamiltonian is 
\begin{equation}
|\psi(h) \rangle = \prod_{k=1}^{K(h)}d_k^\dagger d_{N-k}^\dagger | 0 \rangle,
\end{equation}
 where $|0\rangle$ is the vacuum that satisfies $d_k|0\rangle = 0$ for $k=1,\dots, N$, and $K(h)\leqslant N/2$ is an integer such that $\epsilon_{K(h)} \leqslant 0 < \epsilon_{K(h)+1}$, and 
\begin{equation}
\label{eq:Qgs}
Q |\psi(h) \rangle = \left \lfloor\frac{N}{\pi} \arccos h \right \rfloor |\psi(h) \rangle, \quad |h| \leqslant 1.
\end{equation}
In the large-$N$ limit, the occupation number of the groundstate is thus $ N/(2\pi)\arccos h$.

For $h>1$, the energies $\epsilon_k$ are all strictly positive, so that the groundstate is exactly the vacuum, $|\psi(h>1)\rangle = |0\rangle$. This is a sign of the well-known QPT of the model at $h=1$. 

\subsection{QPT and pure-state fidelity}\label{sec:psf}

Let us apply the notion of pure-state fidelity \cite{zanardi2006ground} to the QPT of the XX chain at $h=1$. We consider the overlap defined in Eq.~\eqref{eq:Fpure},
\begin{equation}
F_{\textrm{pure}}(h,\dd h) = |\langle \psi(h-\dd h) | \psi(h+\dd h)\rangle|,
\end{equation}
with $\dd h > 0$. 

For finite $N$ and $|h|\leqslant 1$, we have $K(h) = \lfloor N  \frac{\arccos h}{2 \pi} \rfloor$, and the occupation number of the groundstate is $2 K(h)$. Moreover, we have $| \psi(1+\dd h)\rangle = |0\rangle$ for all $\dd h>0$. It is possible to choose $\dd h^*=\dd h^*(N,h)$ such that $K(1-\dd h^*)\geqslant 1$, and $K(h-\dd h^*) = K(h+\dd h^*)$ for $|h|<1$. Therefore, because two states with a different occupation number are orthogonal, we have
\begin{equation}
F_{\textrm{pure}}(1,\dd h^*)=0, \qquad F_{\textrm{pure}}(|h|<1,\dd h^*)=1.
\end{equation}

Typically, the variational parameter satisfies $\dd h ^* = \mathcal{O}(N^{-1})$. As we discussed in Sec. \ref{sec:intro}, in the thermodynamic limit $N \to \infty$ all the groundstates are orthogonal and hence pure-state fidelities vanish for all $h$ and $\dd h \neq 0$. For finite sizes, the suppression of the fidelity happens only in the vicinity of the QPT \cite{zanardi2006ground}. 

\section{Thermal fidelities} \label{sec:thermal}

In this section, we investigate symmetry-resolved and total fidelities between thermal density matrices,
\begin{equation}
\label{eq:thermalDM}
\rho(\beta,h) = \frac{\eE^{-\beta H(h)}}{\Tr(\eE^{-\beta H(h)})},
\end{equation}
where $H(h)$ is the Hamiltonian defined in Eq. \eqref{eq:HDiag} and $\beta$ is the inverse temperature. More specifically, we fix a small parameter $\dd h \geqslant 0$ and study the fidelities between $\rho_-~\equiv~\rho(\beta,h-\dd h)$ and $\rho_+\equiv\rho(\beta,h+\dd h)$ as a function of $\beta$ and $h$. We note that both matrices commute: $[\rho_-,\rho_+]=0.$

\subsection{Charged fidelities}

Owing to the diagonal form of the Hamiltonian in Eq. \eqref{eq:HDiag}, we find the following expressions for the charged fidelities, 
\begin{equation}
\label{eq:tildeZnExact}
f_n(\rho_-,\rho_+;\alpha) = \prod_{k=1}^N \frac{1+\eE^{-2\beta n \epsilon_k+\ir \alpha}}{[(1+\eE^{-2\beta n \epsilon_k^-})(1+\eE^{-2 \beta n \epsilon_k^+})]^{1/2}}. 
\end{equation}
This result is consistent with Eq.~\eqref{eq:fnGauss} in the case where $[\rho,\sigma]=0$. However, the diagonal form of the Hamiltonian allowed us to directly evaluate the various traces involved in the definition of $f_n(\rho,\sigma;\alpha)$, see Eq.~\eqref{eq:fnDef} , without using the two-point correlation matrices. 

We also note that the parameter $n$ in Eq. \eqref{eq:tildeZnExact} appears only through the product $\beta n$, which we interpret as an effective inverse temperature. This trivial $n$ dependence stems from the fact that both density matrices commute, as discussed in Sec. \ref{sec:Gaussian}.

To facilitate the forthcoming calculations, we expand $f_n(\rho_-,\rho_+;\alpha)$ at quadratic order in $\alpha$ and $\dd h$. In the large-$N$ limit, we express the sums as integrals and find 
\begin{multline}
\label{eq:tildeZquadr}
\log f_n(\rho_-,\rho_+;\alpha) =\\  \ir \alpha N \mathcal{I}_1- \left( \frac{\alpha^2}{2}+2 (\beta n)^2 \dd h^2\right)N\mathcal{I}_2,
\end{multline}
where we introduced 
\begin{equation}
\label{eq:Ij}
\mathcal{I}_j =\frac{1}{2 \pi} \int_{0}^{2\pi}\dd \theta \frac{\eE^{2\beta n(\cos \theta-h)}}{\left(1+\eE^{2\beta n(\cos \theta-h)}\right)^j}.
\end{equation}

\subsection{Total fidelities} \label{sec:thermalFid}

We start with the computation of the total fidelities $F_n(\rho_-,\rho_+)=f_n(\rho_-,\rho_+;0)$. With Eq.~\eqref{eq:tildeZnExact}, we find
\begin{equation}
\label{eq:FntotEx}
F_n(\rho_-,\rho_+)= \prod_{k=1}^N \frac{1+\eE^{-2\beta n \epsilon_k}}{[(1+\eE^{-2\beta n \epsilon_k^-})(1+\eE^{-2 \beta n \epsilon_k^+})]^{1/2}}. 
\end{equation}
We note that this equation was obtained in Ref. \cite{zanardi2007mixed} for $n=1/2$, where the authors studied the behavior of the Uhlmann-Jozsa fidelity for thermal states close to a QPT in the XY chain. As expected from Eq. \eqref{eq:tildeZnExact}, we observe that $F_n(\rho_-,\rho_+)$ only depends on the index $n$ through the effective temperature $\beta n$. The R\'enyi fidelities thus have the same properties as those observed in Ref. \cite{zanardi2007mixed} for the Uhlmann-Jozsa fidelity. To understand the qualitative behavior of the R\'enyi fidelities, we use Eq.~\eqref{eq:tildeZquadr} and find the following quadratic expansion,
\begin{equation}
F_n(\rho_-,\rho_+) = \eE^{- 2 \dd h^2 (\beta n)^2 N \mathcal{I}_2}.
\label{eq:FntotQd}
\end{equation}

\begin{figure}
\begin{tabular}{l}
\includegraphics[width=0.4\textwidth]{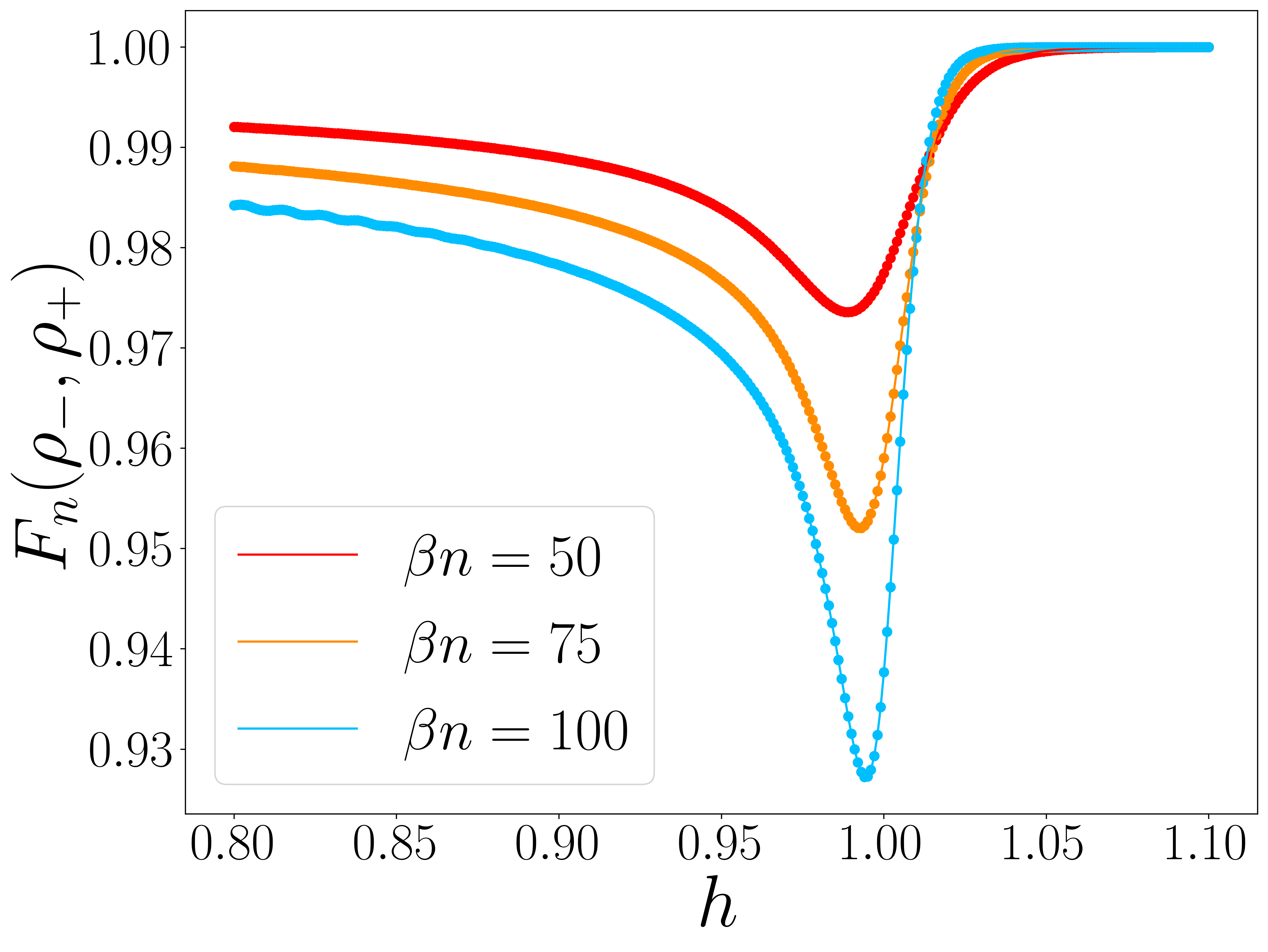} \qquad 
\end{tabular}
\caption{Total fidelities $F_n(\rho_-,\rho_+)$ as a function of $h$ for various values of $\beta n$ with $N=300$ and $\dd h=10^{-3}$. We compare the exact result of Eq.~\eqref{eq:FntotEx} (symbols) with the approximation of Eq.~\eqref{eq:FntotQd} (solid lines) and find excellent agreement. 
} 
\label{Fig:Fn_tildeFn}
\end{figure}

In Fig.~\ref{Fig:Fn_tildeFn}, we plot the total fidelities as a function of $h$ for various values of $\beta n$. We find a very good match between the exact result of Eq.~\eqref{eq:FntotEx} (symbols) and the quadratic approximation of Eq.~\eqref{eq:FntotQd} (solid lines). As expected from Ref. \cite{zanardi2007mixed}, we observe a drop in the total fidelities at $h=1$, which is stronger for smaller effective temperatures. The physical interpretation is that the R\'enyi fidelities are able to detect the zero-temperature QPT of the XX spin chain at $h=1$, even at non-zero temperature. 

Let us discuss some properties of the total fidelities. First, for $\dd h=0$, we have $F_n(\rho_-,\rho_+)=1$ for all values of $h$, and the QPT is no longer detectable via the total R\'enyi fidelities. Second, for $\dd h>0$, the quadratic expression of Eq.~\eqref{eq:FntotQd} allows us to understand that the drop in the fidelities at $h=1$ is entirely due to the behavior of the $\I_2$ integral. For large but finite $\beta n$, $\I_2$ is negligible for all values of $h$, except for a small interval around $h=1$ where it has a sharp peak. 

\subsection{Symmetry-resolved fidelities}
We turn to the computation of the symmetry-resolved fidelities. Combining Eqs. \eqref{eq:FnqInt} and \eqref{eq:tildeZnExact}, we find 
\begin{multline}
\label{eq:FnqDef}
\nnF_n(\rho_-,\rho_+;q) =\\  \int_{-\pi}^{\pi} \frac{\dd \alpha}{2 \pi}\eE^{-\ir \alpha q}\prod_{k=1}^N \frac{1+\eE^{-2\beta n \epsilon_k+\ir \alpha}}{[(1+\eE^{-2\beta n \epsilon_k^-})(1+\eE^{-2 \beta n \epsilon_k^+})]^{1/2}}.
\end{multline}

\subsubsection{Sectors $q=0,1$}
It is possible to obtain exact results for the Fourier transform. For $q=0,1$, it is simply
\begin{equation}
\label{eq:Fq01Ex}
\begin{split}
\nnF_n(\rho_-,\rho_+;0) & =  \prod_{k=1}^N \frac{1}{[(1+\eE^{-2\beta n \epsilon_k^-})(1+\eE^{-2 \beta n \epsilon_k^+})]^{1/2}}, \\
\nnF_n(\rho_-,\rho_+;1) & =   \frac{\sum_{k=1}^N \eE^{-2\beta n \epsilon_k}}{\prod_{k=1}^N[(1+\eE^{-2\beta n \epsilon_k^-})(1+\eE^{-2 \beta n \epsilon_k^+})]^{1/2}}.
\end{split}
\end{equation}
At quadratic order in $\dd h$, we have
\begin{equation}
\label{eq:Fq01Qd}
\begin{split}
\nnF_n(\rho_-,\rho_+;0)  &= \eE^{- N \mathcal{I}_{\log}} \eE^{- 2 \dd h^2 (\beta n)^2  N\mathcal{I}_2 }, \\
\nnF_n(\rho_-,\rho_+;1)  &=N \I_0 \eE^{- N \mathcal{I}_{\log}} \eE^{- 2 \dd h^2 (\beta n)^2  N\mathcal{I}_2 }, \\
\end{split}
\end{equation}
where 
\begin{equation}
\label{eq:IlogEx}
\mathcal{I}_{\log} =\frac{1}{2 \pi} \int_{0}^{2\pi}\dd \theta \log\left(1+\eE^{2\beta n(\cos \theta-h)}\right),
\end{equation}
and $\I_0$ is defined in Eq.~\eqref{eq:Ij}.

\begin{figure*}
\begin{tabular}{l}
\includegraphics[width=0.4\textwidth]{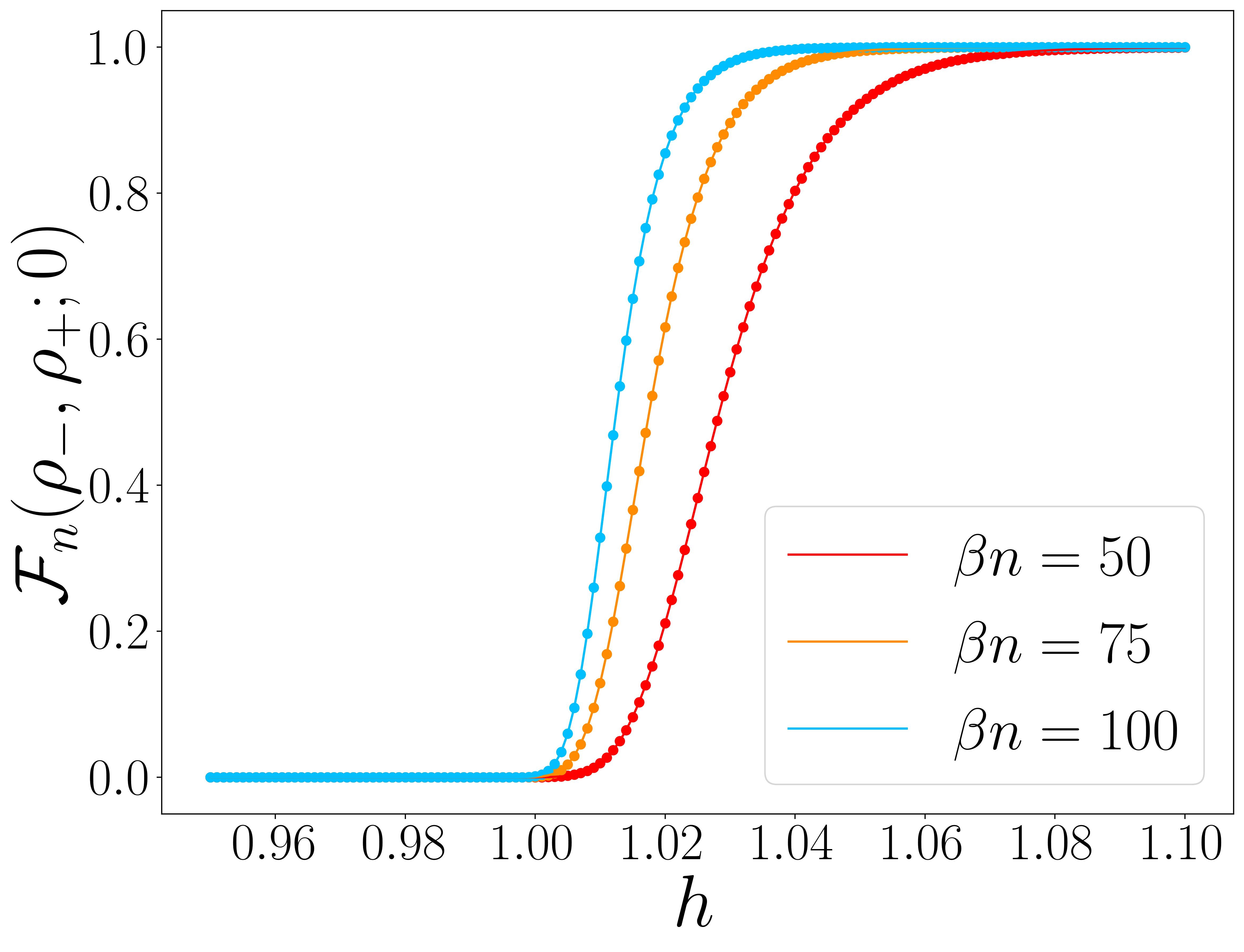} \qquad 
\includegraphics[width=0.4\textwidth]{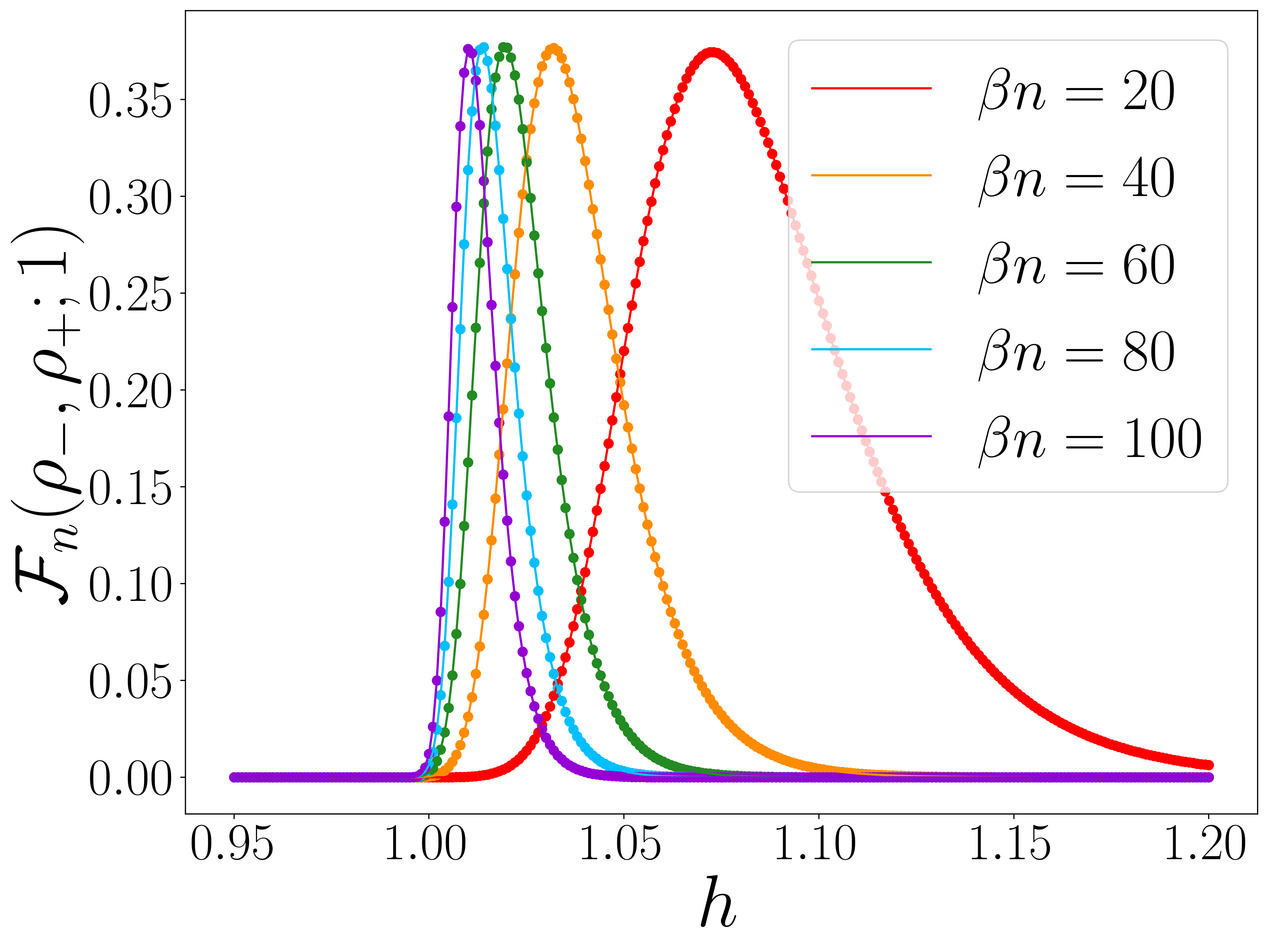}  
\end{tabular}
\caption{Symmetry-resolved fidelities $\nnF_n(\rho_-,\rho_+;q)$ for $q=0$ (left) and $q=1$ (right) as a function of $h$ for various values of $\beta n$ with $N=300$ and $\dd h=10^{-3}$. We compare the exact results of Eq.~\eqref{eq:Fq01Ex} (symbols) with the quadratic approximations of Eq.~\eqref{eq:Fq01Qd} (solid lines) and find excellent agreement.
} 
\label{Fig:tildeFnq01}
\end{figure*}

In Fig.~\ref{Fig:tildeFnq01}, we compare the exact formula of Eq.~\eqref{eq:Fq01Ex} (symbols) for $\nnF_n(\rho_-,\rho_+;0)$ and $\nnF_n(\rho_-,\rho_+;1)$ with their respective quadratic approximation of Eq.~\eqref{eq:Fq01Qd} (solid lines), and find excellent agreement. The behavior of $\nnF_n(\rho_-,\rho_+;0)$ clearly allows one to detect the QPT, as it experiences an abrupt suppression around $h=1$, which is sharper for smaller effective temperatures. In the zero-temperature limit, we expect $\nnF_n(\rho_-,\rho_+;0) \sim \Theta(h-1)$, where $\Theta(x)$ is the Heaviside step function. This is because the integral $\I_{\log}$ in Eq.~\eqref{eq:IlogEx} diverges in the zero-temperature limit for $h<1$ and vanishes for $h\geqslant 1$. Similarly to the total fidelities, $\nnF_n(\rho_-,\rho_+;0)$ thus detects the underlying QPT, even at non-zero temperature.

 However, since the behavior of $\nnF_n(\rho_-,\rho_+;0)$ is mainly dictated by $\I_{\log}$, it is not very sensitive to $\dd h$, and in particular the QPT is still visible in the behavior of $\nnF_n(\rho_-,\rho_+;0)$ for $\dd h=0$. This is in stark contrast with the total fidelities discussed in the previous section. The physical intuition is the following. For $\dd h=0$, $\rho_-=\rho_+=\rho(h,\beta)$, and the symmetry-resolved fidelities $\nnF_n(\rho_+,\rho_-;q)$ are essentially governed by the probabilities $p_{\rho(h,\beta)}(q)$. While the probabilities sum to one for all values of $h$, they undergo a strong reorganization near $h=1$. The QPT thus expresses itself in the probability distribution $\{p_{\rho(h,\beta)}(q)\}_q$. The total fidelities are not sensitive to this reorganization, because they involve a sum over all probabilities, whereas the symmetry-resolved ones are. We conclude that the symmetry-resolved fidelities probe the structure of the density matrices and detect reorganization of a state as it undergoes a QPT even when this reorganization does not affect the total fidelities. 

\subsubsection{Arbitrary $q$}
\begin{figure}
\begin{tabular}{l}
\includegraphics[width=0.4\textwidth]{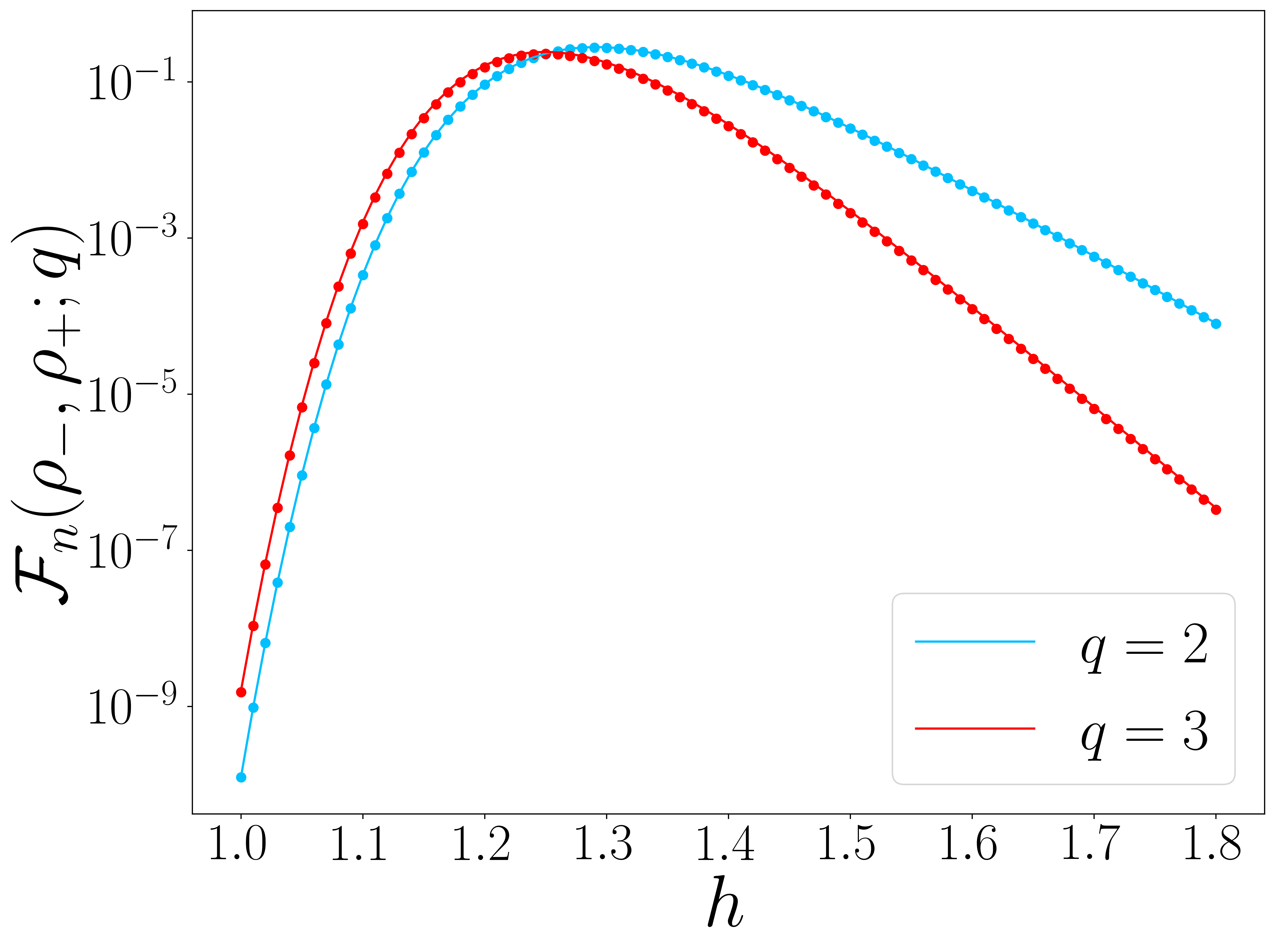}   
\end{tabular}
\caption{Symmetry-resolved fidelities $\nnF_n(\rho_-,\rho_+;q)$ for $q=2,3$ in log scale as a function of $h$ with $\beta n=5$, $N=300$, and $\dd h=10^{-3}$. The symbols are obtained by numerical evaluation of the Fourier transform in Eq.~\eqref{eq:FnqDef} and the solid lines are given by Eq.~\eqref{eq:FnqEx}, where $\nnF_n(\rho_-,\rho_+;0)$ is replaced by its quadratic approximation of Eq.~\eqref{eq:Fq01Qd}. We find very good agreement. 
} 
\label{Fig:tildeFnq23}
\end{figure}

For arbitrary $q$ and in the large-$N$ limit, the symmetry-resolved fidelities read
\begin{equation}
\label{eq:FnqEx}
\nnF_n(\rho_-,\rho_+;q) =N^q \mathcal{J}_q \ \nnF_n(\rho_-,\rho_+;0), 
\end{equation}
with
\begin{multline}
\mathcal{J}_q =\\ \frac{1}{(2 \pi)^q}\int_{0}^{2\pi}\dd \theta_1 \int_{\theta_1}^{2\pi}\dd \theta_2 \cdots \int_{\theta_{q-1}}^{2\pi}\dd \theta_q \eE^{2\beta n\sum_{j=1}^q (\cos \theta_j-h)}
\end{multline}
and $\mathcal{J}_0 \equiv 1$. We also note that $\mathcal{J}_1=\I_0$, so Eq.~\eqref{eq:FnqEx} is compatible with Eq.~\eqref{eq:Fq01Qd} for $q=0,1$. We verify Eq.~\eqref{eq:FnqEx} in Fig.~\ref{Fig:tildeFnq23} for $q=2,3$. In particular, we compare Eq.~\eqref{eq:FnqDef} (symbols) with Eq.~\eqref{eq:FnqEx} where $\nnF_n(\rho_-,\rho_+;0)$ is replaced by its quadratic approximation of Eq.~\eqref{eq:Fq01Qd} (solid lines). We find very good agreement.

For $q>1$, even though Eq.~\eqref{eq:FnqEx} is exact in the large-$N$ limit, it is not suitable to understand the qualitative physical behavior of the symmetry-resolved fidelities. To do so, we use the quadratic approximation of Eq.~\eqref{eq:tildeZquadr} for the charged fidelities and perform the integral in Eq.~\eqref{eq:FnqInt}. We find
\begin{multline}
\label{eq:FnqQdr}
\nnF_n(\rho_-,\rho_+;q) =\\  \eE^{-2 \dd h^2 (\beta n )^2 N \mathcal{I}_2} 
  \eE^{- \frac{\left(q- N \mathcal{I}_1\right)^2 }{2N \mathcal{I}_2}}\sqrt{\frac{1}{2\pi N \mathcal{I}_2}}.
\end{multline}

\begin{figure*}
\begin{tabular}{l}
\includegraphics[width=0.4\textwidth]{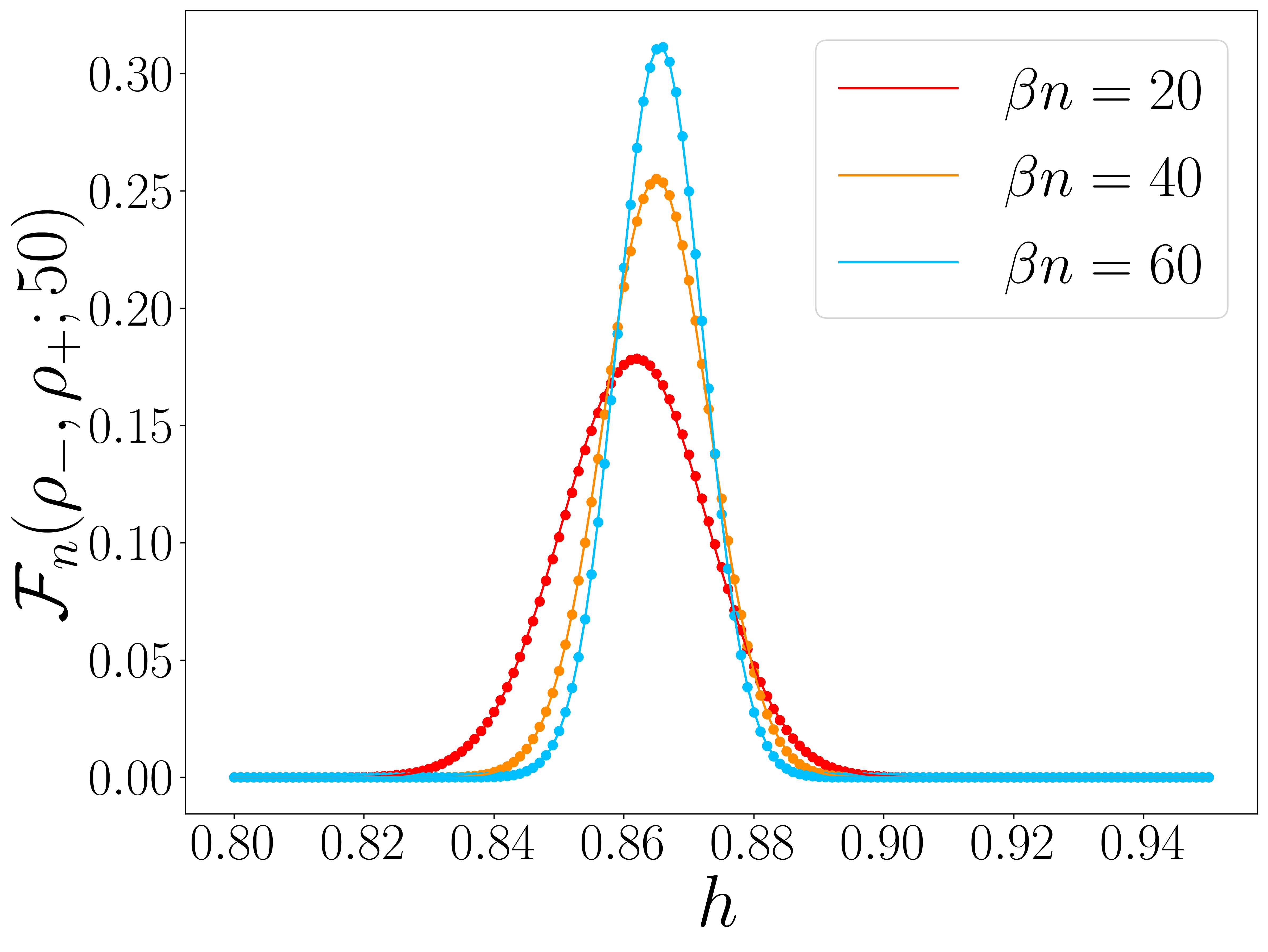} \qquad
\includegraphics[width=0.4\textwidth]{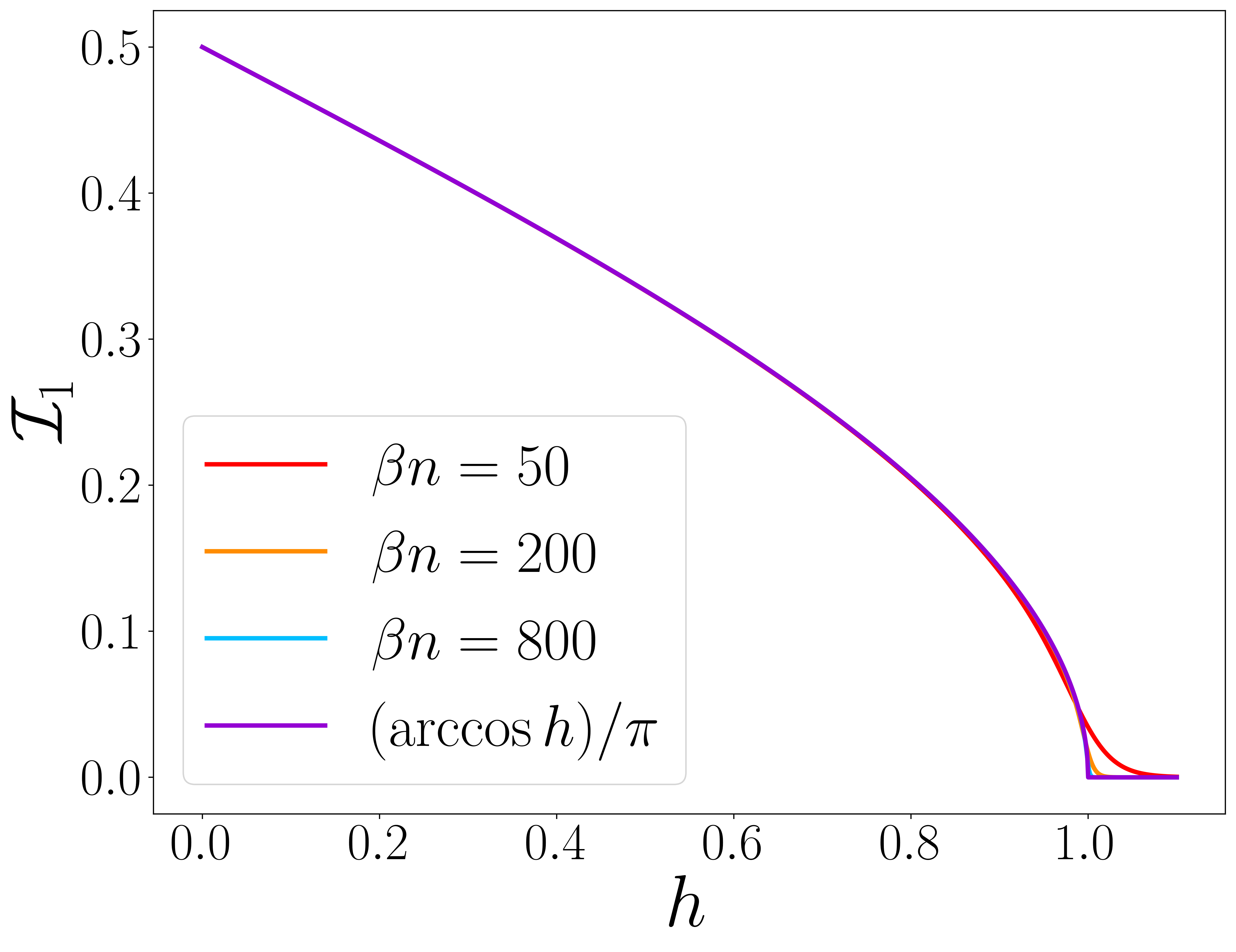}   
\end{tabular}
\caption{\textit{Left}: Symmetry-resolved fidelities $\nnF_n(\rho_-,\rho_+;50)$ as a function of $h$ for various values of $\beta n$ with $N=300$ and $\dd h=10^{-3}$. We compare Eq.~\eqref{eq:FnqDef} (symbols) with the quadratic approximation of Eq.~\eqref{eq:FnqQdr} (solid lines), and find excellent agreement. \textit{Right}: Comparison between the integral $\I_1$ defined in Eq.~\eqref{eq:Ij} for various values of $\beta n$ and the function $(\arccos h)/\pi$.
} 
\label{Fig:tildeFnq}
\end{figure*}

In the left panel of Fig.~\ref{Fig:tildeFnq}, we compare Eq.~\eqref{eq:FnqDef} (symbols) with the quadratic approximation of Eq.~\eqref{eq:FnqQdr} (solid lines) for $\nnF_n(\rho_-,\rho_+;q)$ as a function of $h$ for various values of $\beta n$, and find excellent agreement. We note that the curves are narrower for larger $\beta n$, and their maximum is located at a value $h^*$ that depends on $\beta n$. From Eq.~\eqref{eq:FnqQdr}, it appears that $h^*$ is such that 
\begin{equation}
\label{eq:qNI}
q=N\I_1(h^*),
\end{equation}
where $\I_1$ is defined in Eq.~\eqref{eq:Ij}, and we emphasize its dependence on the magnetic field. This is not an explicit equation for $h^*$, since it appears in the kernel of the integral $\I_1$. However, we verified numerically that all the peaks we observe in the right panel of Fig.~\ref{Fig:tildeFnq01} and the left panel of Fig.~\ref{Fig:tildeFnq} are precisely located at $h^*$ such that Eq.~\eqref{eq:qNI} is verified. 

In the zero-temperature limit, we expect the peaks to be located at $h^*$ such that $q=(N \arccos h^*)/\pi$ is the occupation number of the groundstate, see Eq.~\eqref{eq:Qgs}. We prove this physical intuition with the following asymptotic analysis of $\I_1$ in the limit $\beta n \to \infty$. The integrand of $\I_1$ is $(1+ \eE^{-2\beta n(\cos \theta-h)})^{-1}$. In the zero-temperature limit, this quantity tends to zero for $\cos \theta<h$, and to one for $\cos \theta>h$. The integral $\I_1$ is thus proportional to the length of the interval in $[0,2\pi]$ on which $\cos \theta>h$. A direct calculation gives 
\begin{equation}
\lim_{\beta n\to \infty} \I_1 = \frac{\arccos h}{\pi}, \quad h\leqslant 1,
\end{equation}
and the integral vanishes for $h>1$. We illustrate this limit in the right panel of Fig.~\ref{Fig:tildeFnq}, where we compare the integral $\I_1$ for various values of $\beta n$ with the asymptotic value $(\arccos h)/ \pi$. 

The fact that the leading charge sector is located at $q= N \I_1$, which is different from the groundstate occupation number, but converges to that value in the zero-temperature limit, is a surprising result of this paper. Physically, $N \I_1$ is the average value of the charge in the thermal state $\rho(\beta,h)$. The average is defined as 
\begin{equation}
\langle Q \rangle _{\beta,h} =\Tr(\rho(\beta,h)Q),
\end{equation}
and we have
\begin{equation}
\begin{split}
\langle Q \rangle _{\beta,h}&=-\ir (\partial_\alpha\log f_{1/2}(\rho_-, \rho_+;\alpha))|_{\alpha=0,\dd h=0} \\
&= N \I_1,
\end{split}
\end{equation}
where we used the quadratic expansion in Eq.~\eqref{eq:tildeZquadr}. 

\subsection{Total fidelities from symmetry-resolved ones}

As a consistency check, we verify that the sum over all sectors of the symmetry-resolved fidelities yields the total ones. Since we work in the large-$N$ limit, we transform the sum over $q$ into an integral, and use the quadratic approximation of Eq.~\eqref{eq:FnqQdr} for the symmetry-resolved fidelities. We find
\begin{multline}
\sum_{q=0}^N \nnF_n(\rho_-,\rho_+;q) \simeq \\  \eE^{-2 \dd h^2 (\beta n )^2 N \mathcal{I}_2}  \int_0^N \dd q \
  \eE^{- \frac{\left(q- N \mathcal{I}_1\right)^2 }{2N \mathcal{I}_2}}\sqrt{\frac{1}{2\pi N \mathcal{I}_2}}. 
\end{multline}
For large $N$, the Gaussian integral on the right-hand side yields $ \eE^{-2 \dd h^2 (\beta n )^2 N \mathcal{I}_2} $, which is exactly the Gaussian approximation of the total fidelities, see Eq.~\eqref{eq:FntotQd}. 

\section{Reduced fidelities} \label{sec:red}

In this section, we study the total and symmetry-resolved reduced fidelities in the XX spin chain. We consider a bipartite chain $A \cup B $ of length $N$, where $A=\{1,\dots,N_A\}$ is a segment of length $N_A < N$. The whole system is at zero temperature in the pure state $|\psi(h)\rangle$, and we focus on the reduced density matrix $\rho_A(h) = \Tr_B (|\psi(h) \rangle \langle \psi(h) |)$. More specifically, we fix a small parameter $\dd h \geqslant 0$ and investigate the fidelities between the reduced density matrices $\rho_{A-} \equiv \rho_A(h-\dd h) $ and $\rho_{A+} \equiv \rho_A(h+\dd h)$ as a function of $h$.  

Because the XX Hamiltonian is quadratic in terms of fermion operators, the reduced density matrices $\rho_{A\pm}$ are fermionic Gaussian operators and we use the results of Sec.~\ref{sec:Gaussian}. In the large-$N$ limit, the two-point correlation matrix $[C_A(h)]_{x,x'} = \Tr_A(\rho_A(h) c_x^\dagger c_{x'})$, $x,x' \in A$, has entries
 \begin{equation}
 [C_A(h)]_{x,x'}= \frac{\sin[(\arccos h)(x-x')]}{\pi (x-x')}, \quad |h| \leqslant 1,
 \end{equation}
 and $ [C_A(h)]_{x,x'}=0$ for $h>1$. From the matrices $C_A(h\pm \dd h)$, we construct the corresponding matrices $ J_A(h\pm \dd h) = 2 C_A(h\pm \dd h)- \id$, and $J_\bullet$ as in Eq.~\eqref{eq:Jbull}. We diagonalize these matrices numerically, and insert the results in Eq.~\eqref{eq:fnGauss} to obtain the charged fidelities. Finally, we evaluate the result at $\alpha=0$ to obtain the total fidelities $F_n(\rho_{A-},\rho_{A+})$, or compute the Fourier transform as in Eq.~\eqref{eq:FnqInt} to evaluate the symmetry-resolved fidelities $\nnF_n(\rho_{A-},\rho_{A+};q)$ . 
 
\subsection{Total fidelities}
 
\begin{figure*}
\begin{tabular}{l}
\includegraphics[width=0.4\textwidth]{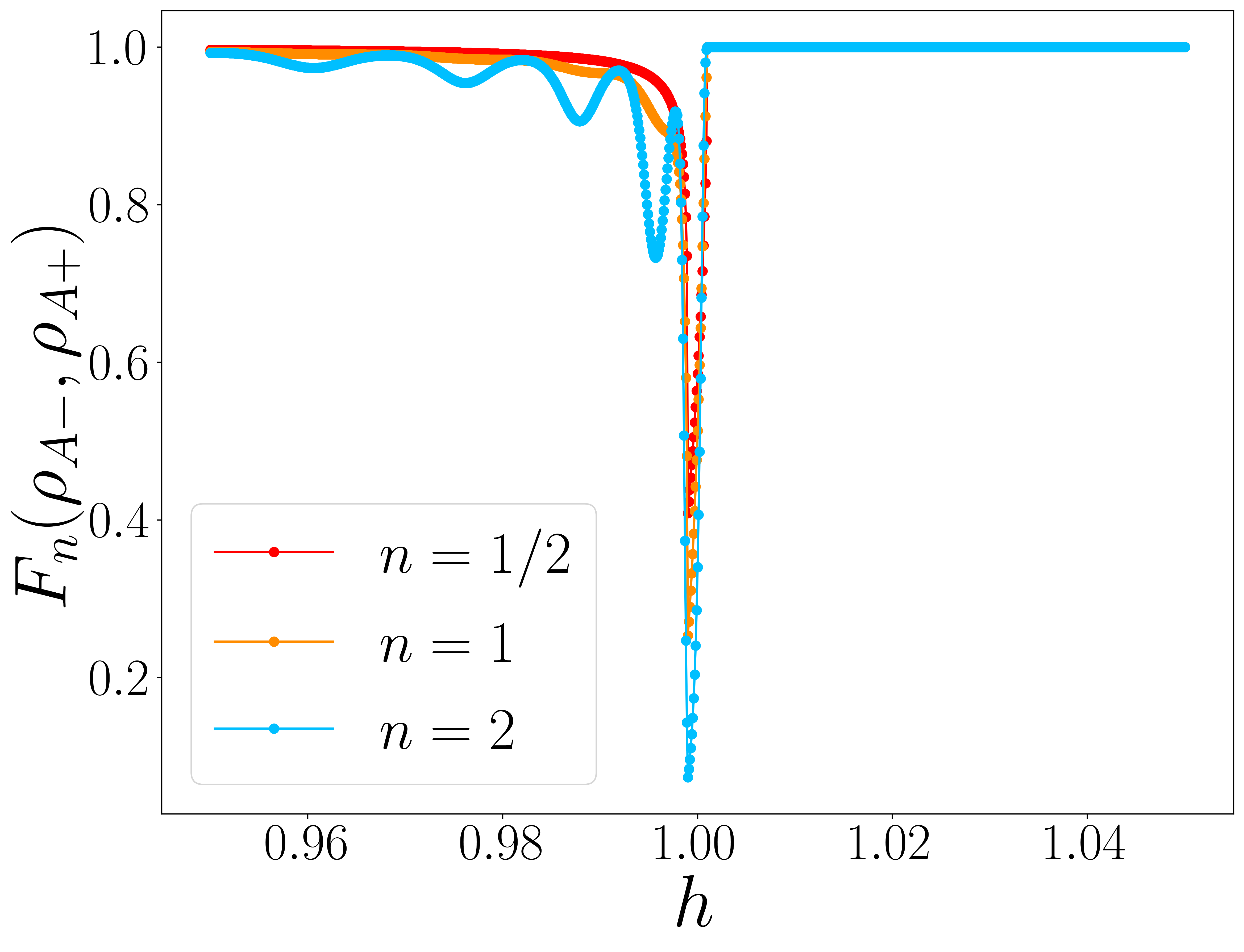}  \qquad 
\includegraphics[width=0.4\textwidth]{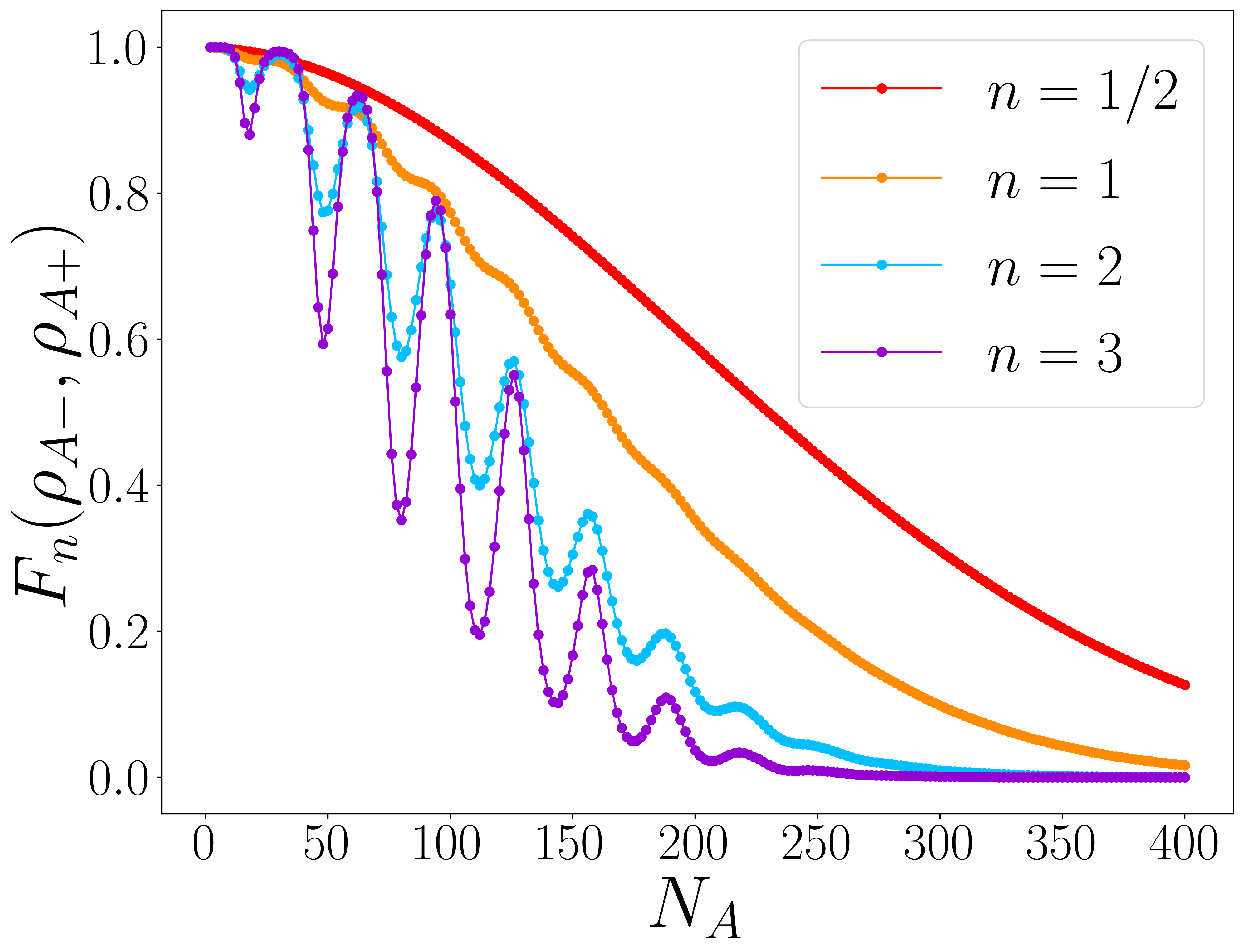} 
\end{tabular}
\caption{\textit{Left}: Total fidelities $F_n(\rho_{A-},\rho_{A+})$ with $N_A=50$ and $\dd h=10^{-3}$ as a function of $h$ for various values of $n$. \textit{Right}: Total fidelities $F_n(\rho_{A-},\rho_{A+})$ with $h=0.995$ and $\dd h=10^{-3}$ as a function of $N_A$ for various values of $n$. In all panels, the symbols are obtained from the numerical diagonalization of the correlation matrices and Eq.~\eqref{eq:fnGauss} at $\alpha=0$. The solid lines just connect the symbols as a guide to the eye, and do not reflect any analytical prediction or conjecture.} 
\label{Fig:FnRedTot}
\end{figure*}

Let us discuss our results for the total fidelities. We display our numerical results for $F_n(\rho_{A-},\rho_{A+})$ with $N_A=50$ and $\dd h=10^{-3}$ as a function of $h$ for various values of $n$ in the left panel of Fig.~\ref{Fig:FnRedTot}. We observe a clear suppression of the fidelities at $h=1$, and this suppression is sharper for increasing values of $n$. Even though the whole system is at zero-temperature, the parameter~$n$ plays the role of an effective inverse temperature in system $A$, similarly as for the thermal fidelities in Sec. \ref{sec:thermalFid}. We also observe that the reduced fidelities oscillate in $h$ with an amplitude that increases with $n$. 

From these results, we conclude that the reduced fidelities are able to detect the QPT at $h=1$. We stress that the system is in the thermodynamic limit $N \to \infty$, and hence the reduced fidelities can detect QPTs in that limit, provided that the subsystem $A$ is finite. This is in stark contrast with pure-state fidelities that detect QPTs only for finite sizes, because all overlaps vanish in the thermodynamic limit, see Ref. \cite{zanardi2006ground} and Sec. \ref{sec:psf}.

In the right panel of Fig.~\ref{Fig:FnRedTot}, we show the total fidelities at $h=0.995$ with $\dd h=10^{-3}$ as a function of $N_A$ for various values of~$n$. We observe oscillations in $N_A$, whose amplitude increases with~$n$ and which confirms that the suppression of the fidelity is more important for larger values of~$n$. We find that the oscillations have a frequency close to $2 \arccos h$ for $h \lesssim 1$. Similar parity oscillations of the entanglement entropy in the XX chain are known to depend on universal quantities of the underlying CFT \cite{calabrese2010parity,calabrese2010universal,fagotti2011universal,bons,mrc-20,berthiere2022entanglement}, and model-dependent oscillations were recently discovered in the context of inhomogeneous spin chains \cite{bernard2022entanglement}. It is an intriguing question to understand if oscillations in the reduced fidelities also bear relevant physical information about the underlying model, but we leave this issue for forthcoming investigations. 

\subsection{Symmetry-resolved fidelities}

\begin{figure*}
\begin{tabular}{l}
\includegraphics[width=0.4\textwidth]{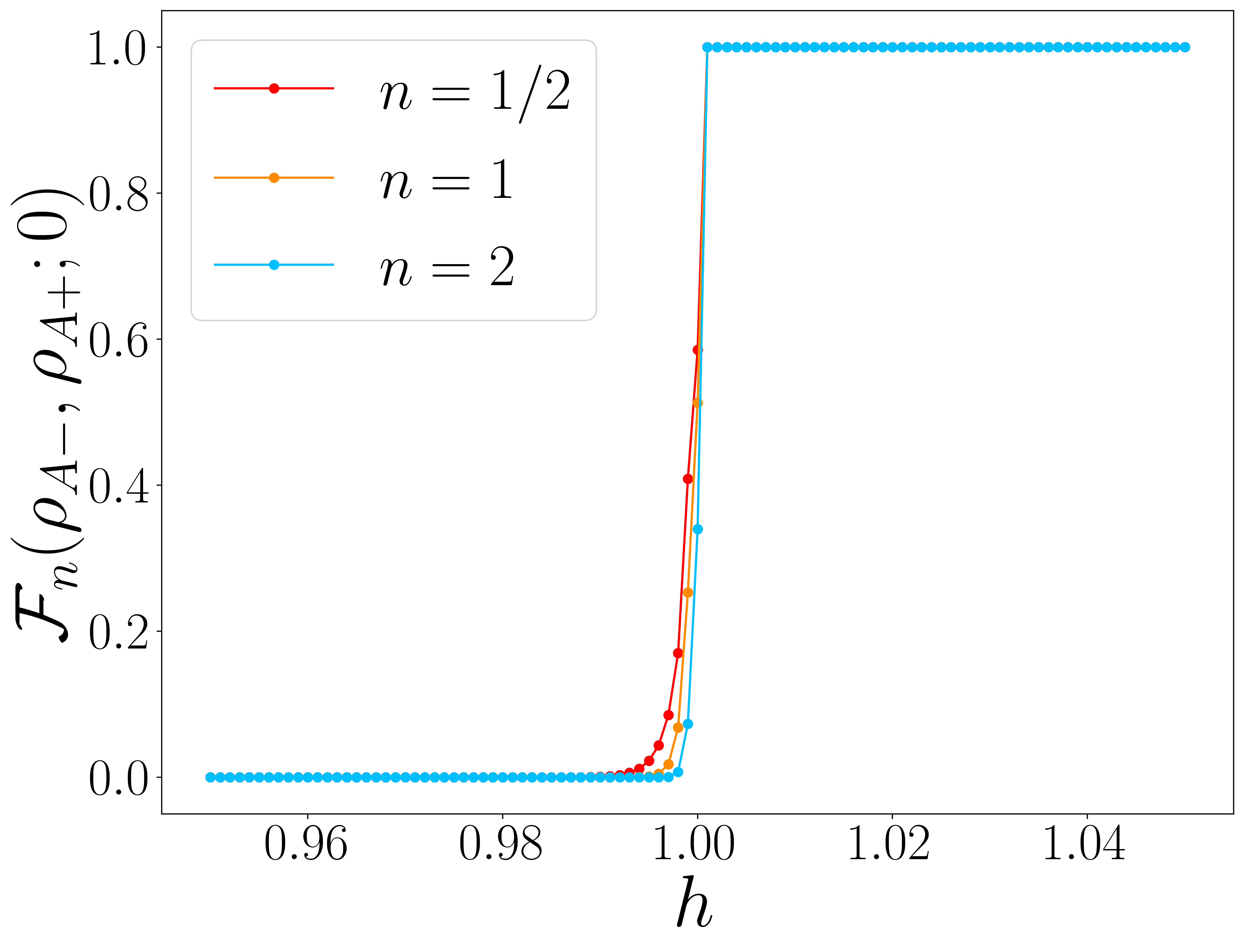} \qquad
\includegraphics[width=0.4\textwidth]{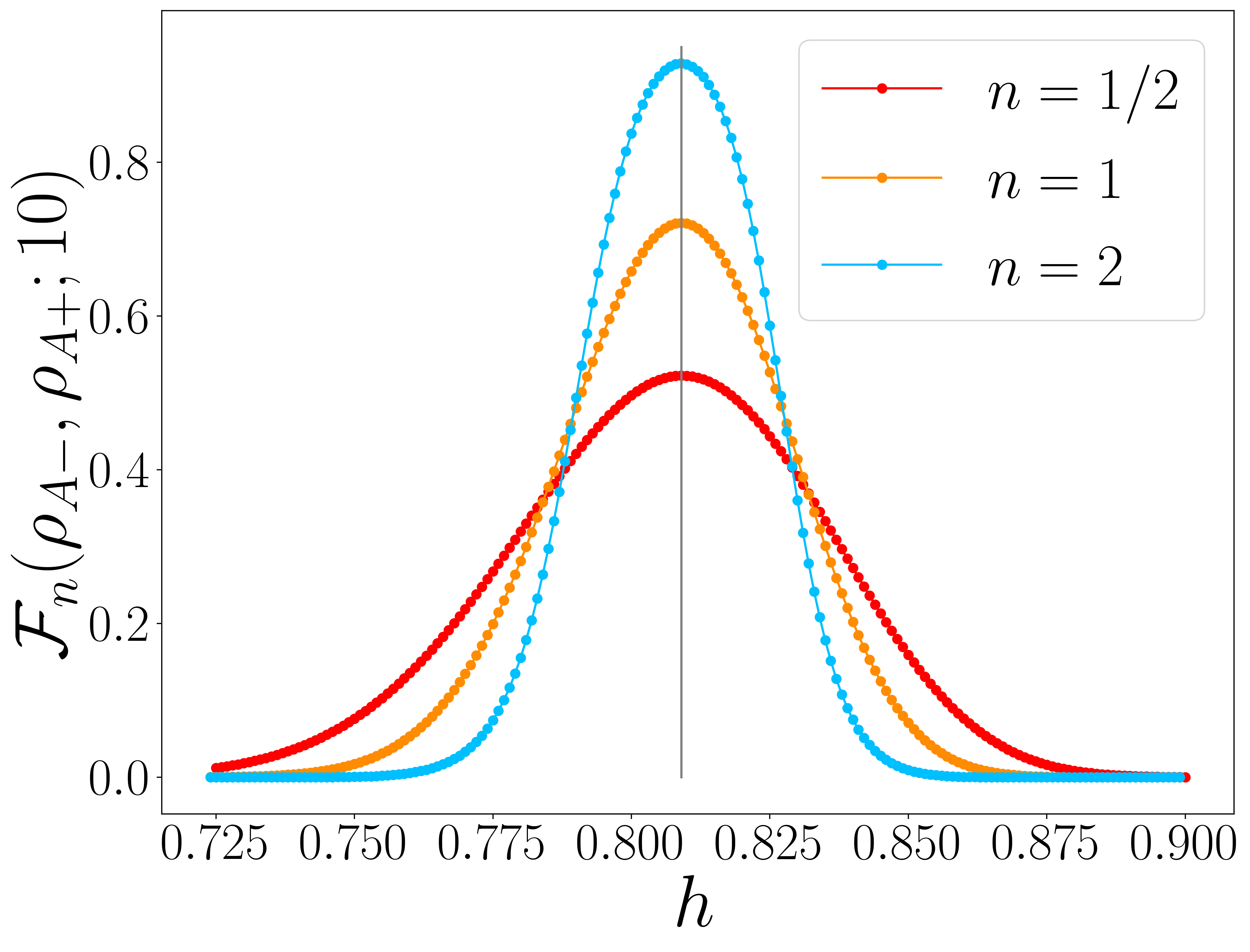} 
\end{tabular}
\caption{Symmetry-resolved fidelities $\nnF_n(\rho_{A-},\rho_{A+};q)$ for $q=0$ (left) and $q=10$ (right) with $N_A=50$ and $\dd h=10^{-3}$ as a function of $h$ for various values of $n$. In the right panel, the grey line is at position $h=\cos(\pi/5)$. The symbols are obtained from the numerical diagonalization of the correlation matrices and the Fourier transform of Eq.~\eqref{eq:fnGauss}. The solid lines just connect the symbols as a guide to the eye, and do not reflect any analytical prediction or conjecture.
} 
\label{Fig:FnqRed}
\end{figure*}

We show our results for the symmetry-resolved fidelities in Fig.~\ref{Fig:FnqRed}. For $q=0$, we have the same qualitative behavior as for thermal fidelities, namely, $\nnF_n(\rho_{A-},\rho_{A+};0)$ drops abruptly at $h=1$, and tends to $\Theta(h-1)$ in the limit $n \to \infty$. Moreover, this behavior holds irrespective of $\dd h$. This highlights the fact that symmetry-resolved fidelities probe the inner structure of the states, and detect the QPT through the reorganization of the charge sectors at the critical point, even though this reorganization does not affect the total fidelities. 

For $q>0$, we observe peaks that are narrower for larger $n$, which is compatible with the interpretation of~$n$ as an effective inverse temperature. However, in contrast with thermal fidelities, the peaks are all located at $h=\cos(q \pi/N_A)$; see the right panel of Fig.~\ref{Fig:FnqRed}. In particular, for fixed $h$, the maximal contribution arises from the sector at $q=(N_A/\pi)\arccos h$. Physically, this value of $q$ is the average value of the charge in system $A$, and since there are no thermal fluctuations we do not observe temperature-dependent peaks, as opposed to thermal fidelities. As a consistency check, we also numerically verified Eq.~\eqref{eq:Ftotq}.

\section{Conclusion} \label{sec:ccl}

\subsection{Summary}

In this paper, we initiated the investigation of symmetry-resolved quantum fidelities in quantum many-body systems. First, we introduced a family of quantum fidelities $F_n(\rho,\sigma)$, that we call R\'enyi fidelities. We showed that these quantities satisfy natural generalizations of the axioms satisfied by the Uhlmann-Jozsa fidelity and, in particular, they reduce to the latter for $n=1/2$. Second, we defined the (non-normalized) symmetry-resolved R\'enyi fidelities $\nnF_n(\rho,\sigma;q)$ and expressed them as a Fourier transform of the charged fidelities $f_n(\rho,\sigma;\alpha)$. In particular, we derived an exact expression for the charged fidelities of fermionic Gaussian states in terms of the associated two-point correlation matrices. This formula, Eq.~\eqref{eq:fnGauss}, is a main result of this paper because not only does it allow one to compute symmetry-resolved fidelities, it also provides an expression for the total R\'enyi fidelities of fermionic Gaussian states, which was lacking in the literature. We believe that this result will allow for a much-needed systematic investigation of R\'enyi fidelities in the context of quantum many-body systems, both in and out of equilibrium. 

We investigated the total and symmetry-resolved fidelities in the vicinity of the QPT at $h=1$ in the XX spin chain. After a brief review of the pure-state fidelities, we devoted our attention to the thermal and reduced fidelities. In the context of thermal fidelities, we used the diagonal form of the Hamiltonian to derive a variety of exact results and quadratic approximations that we verified with extensive numerical investigations. For the total fidelities, we derived an exact expression which generalizes the one of Ref. \cite{zanardi2007mixed} to arbitrary $n$. We found that the quantities only depend on $n$ through the product $\beta n$, where $\beta$ is the inverse temperature. As in Ref.~\cite{zanardi2007mixed}, we concluded that the thermal R\'enyi fidelities are able to detect the zero-temperature QPT of the XX Hamiltonian, even when the system is at finite temperature. In addition, we understood this result qualitatively with the Gaussian approximation of Eq.~\eqref{eq:FntotQd} and, in particular, we noted the importance of the integral $\I_2$ in the behavior of the fidelities close to $h=1$. 

We then investigated the symmetry-resolved thermal fidelities. For $q=0$, we observed that $\nnF(\rho_-,\rho_+;0)$ is able to detect the QPT, even for $\dd h=0$, which is in stark contrast with the total fidelities. Physically, this means that the symmetry-resolved fidelities are sensitive to the inner structure of the states, and to the strong reorganization that takes place at the QPT, even though this reorganization is not detectable with the total fidelities. For $q>0$, the symmetry-resolved fidelities have peaks located at $q= N \I_1$ with a width that decreases with $\beta n$. In the zero-temperature limit, the position of the maxima converges to the groundstate occupation number $q=(N/\pi) \arccos h$, but it varies at finite temperature because of thermal fluctuations. We showed that the charge $q= N \I_1$ is exactly the average value of the charge in the thermal state $\rho(\beta,h)$.

We turned to the reduced fidelities in the case where $A$ is a finite chain embedded in an infinite one. Using our formulas for Gaussian states and numerical diagonalization of the correlation matrices, we showed that the total fidelities are able to detect the QPT at $h=1$. This shows that reduced fidelities, unlike pure-state fidelities, can be used to detect QPTs in the thermodynamic limit. To the best of our knowledge, this is the first time that a reduced fidelity is computed in the case where subsystem $A$ is a genuine many-body system. Moreover, we observed oscillations in the total reduced fidelities, both as a function of $h$ with fixed $N_A$, and the contrary. The amplitude of the oscillations increases with $n$. For the oscillations with $N_A$ at fixed $h$, we verified numerically that the oscillations have a frequency close to $2 \arccos h$ for $h \lesssim 1$, which is the same frequency as the parity oscillations of the entanglement entropy \cite{calabrese2010parity,calabrese2010universal,fagotti2011universal}. However, we did not conjecture the exact formula for the oscillations, and leave this issue for forthcoming investigations. 

To conclude, we investigated the symmetry-resolved reduced fidelities. For $q=0$, we observed the same qualitative features as for the thermal fidelities, namely, $\nnF_n(\rho_A-, \rho_A+;0)$ is able to detect the QPT at $h=1$, irrespective of $\dd h$. This thus provides a second example where the symmetry-resolved fidelities can detect the QPT through its action on the inner structure of the states, which does not affect the total fidelities. This behavior is a main result of this paper, and we expect it to be a general feature of symmetry-resolved fidelities. For $q>0$, the symmetry-resolved fidelities have peaks with a width that decreases with $n$. However, in contrast with thermal fidelities, all peaks reach their maxima at $h=\cos(q \pi/N_A)$, which implies that for fixed $h$, the leading sector is $q=(N_A/\pi)\arccos h$. This value of the charge is the average charge of system $A$, and there are no thermal fluctuations.  

\subsection{Outlook}

We finish with a series of open questions that would be worth investigating in the future. First, it would be interesting to study symmetry-resolved fidelities in the vicinity of other phase transitions, including topological ones. It would also be important to generalize our results in interacting critical systems, such as the XXZ spin chain. Second, the so-called Loschmidt echo \cite{peres1984stability,QSLZS06,gorin2006dynamics} generalizes fidelities in non-equilibrium systems. It would certainly be interesting to investigate the symmetry resolution of this quantity to probe how each symmetry sector contributes out of equilibrium. It is known that Loschmidt echo \cite{stephan2011local}, as well as symmetry-resolved entropies \cite{PBC21,PBC21bis}, evolve according to the quasiparticle picture \cite{calabrese2005evolution,alba2017entanglement} out of equilibrium, and generalizing these results to symmetry-resolved fidelities seems promising. A related issue is to investigate symmetry-resolved fidelities during a dynamical phase transition~\cite{heyl2018dynamical}, including recently discovered third-order ones \cite{perez2022dynamical}. Another important point, which we have not addressed here, is to study symmetry-resolved fidelities in the context of CFT. In particular, the methods developed in Refs.~\cite{zhang2019subsystem,cc-21} could allow us to investigate the fidelities for excited states, but we expect it to be a hard problem. We also mention that, as a by-product, our results for thermal charged fidelities provide a method to test recent predictions made for thermal symmetry-resolved entanglement in CFT \cite{ghasemi2022universal,tang2022thermal}.

On the analytical side, it would be important to understand the behavior of reduced fidelities from the asymptotic methods developed to compute entanglement entropies in the XX chain \cite{jin2004,its2005entanglement,calabrese2010universal,fagotti2011universal}. However, we expect this to be a hard problem, because the matrix $J_\bullet$ typically does not commute with the correlation matrices. It is a similar technical issue that makes exact asymptotic calculations of fermionic negativities a notoriously hard problem. Exact methods to compute negativities out of equilibrium have recently been discovered \cite{alba2022logarithmic,FG22,bertini2022entanglement}, and we expect that they will apply to reduced fidelities. Another important question regarding reduced fidelities is to understand their oscillations and, in particular, discover if they contain universal quantities, similarly to the entanglement entropies. 

Another problem is to investigate fidelities in the context of inhomogeneous systems. Entanglement measures in inhomogeneous systems have attracted a lot of attention \cite{ramirez2015entanglement,viti2016inhomogeneous,DSVC17,rodriguez2017more,Crampe:2019upj,Crampe:2021,scopa2022exact,bernard2022entanglement}, and we hope that investigating fidelities in the same context will deepen our understanding of these physical systems. On general grounds, we think that most research questions raised in the context of entanglement measures are also worth investigating through the lenses of fidelities. 

\vspace{.4cm}
\section*{Acknowledgments}  The author holds a CRM-ISM postdoctoral fellowship and acknowledges support from the Mathematical Physics Laboratory of the CRM. He warmly thanks Pasquale Calabrese and Christian Hagendorf for fruitful discussions and encouragements, and Cl\'ement Berthiere, Rufus Boyack, Alexandre Lazarescu, Sandrine Brasseur, Riccarda Bonsignori and Sara Murciano for stimulating discussions and comments on the manuscript. 

%

\providecommand{\href}[2]{#2}\begingroup\raggedright\endgroup

\end{document}